\documentclass[12pt]{article}
\usepackage{a4}
\usepackage{latexsym} 
\usepackage{amssymb}  
\usepackage{graphicx}
\usepackage{epsfig}
\usepackage{amsmath}
\usepackage{amsfonts}
\usepackage{color}
\newcounter{zyxabstract}     
\newcounter{zyxrefers}        

\newcommand{\newabstract}
{\newpage\stepcounter{zyxabstract}\setcounter{equation}{0}
\setcounter{footnote}{0}}

\newcommand{\rlabel}[1]{\label{zyx\arabic{zyxabstract}#1}}
\newcommand{\rref}[1]{\ref{zyx\arabic{zyxabstract}#1}}

\renewenvironment{thebibliography}[1] 
{\section*{References}\setcounter{zyxrefers}{0}
\begin{list}
{[\arabic{zyxrefers}]}
{\usecounter{zyxrefers}\setlength{\parindent}{0cm}\setlength{\itemsep}{0cm}}

}
{\end{list}}
{\section*{References}\setcounter{zyxrefers}{0}
\begin{list}{[\arabic{zyxrefers}]}
{\usecounter{zyxrefers}\setlength{\parindent}{0cm}\setlength{\itemsep}{-1.5mm}}}
{\end{list}}

\renewcommand{\bibitem}[1]{\item\rlabel{y#1}}
\renewcommand{\cite}[1]{[\rref{y#1}]}      

\def\e{\kern+.5ex\lower.42ex\hbox{$\scriptstyle \iota$}\kern-1.10ex e}

\begin{document}
\begin{titlepage}

\begin{flushleft} 
\fbox{\includegraphics*[height=2.2cm]{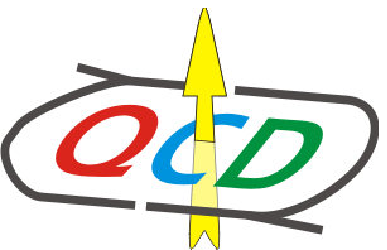}}~~
\includegraphics*[height=2.1cm]{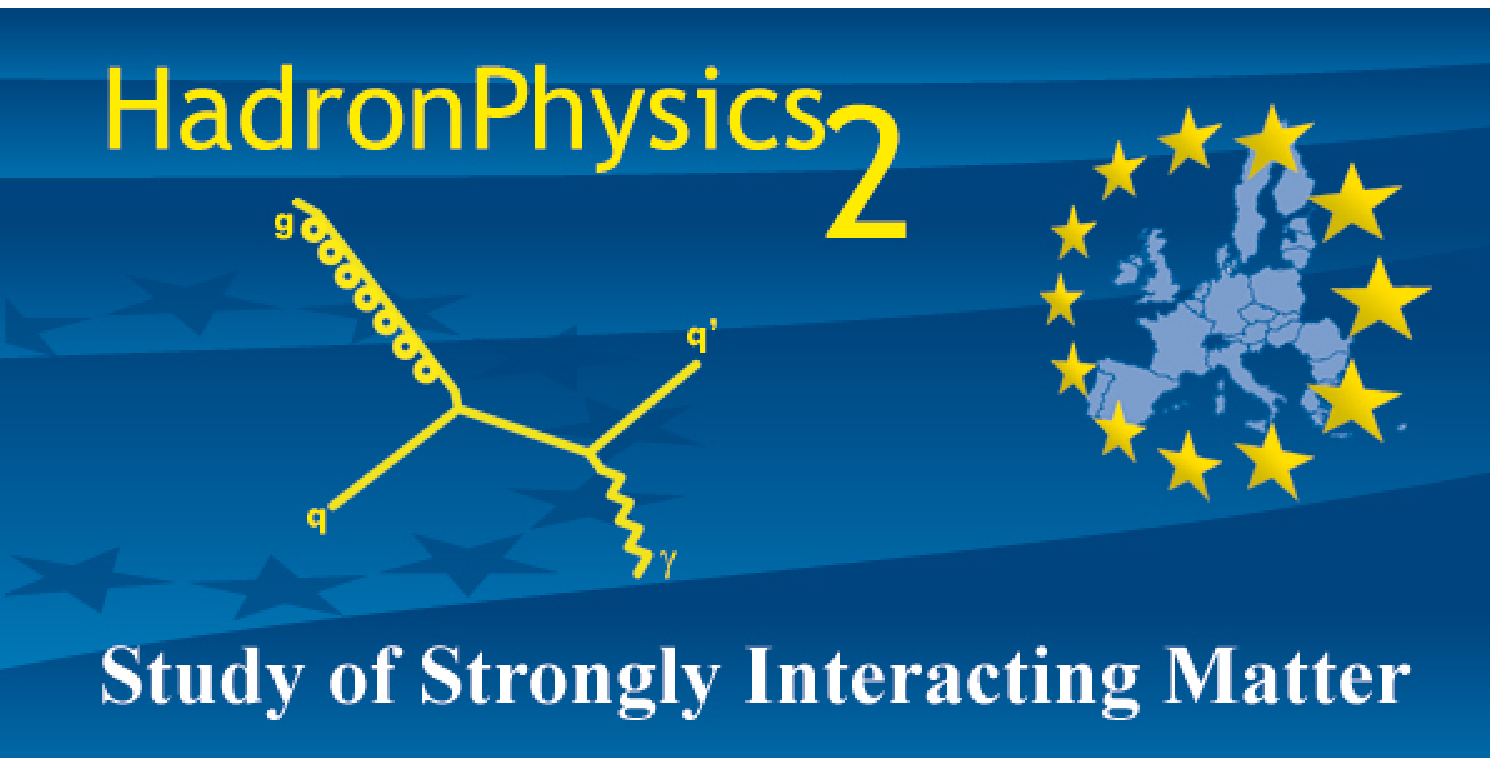}
\end{flushleft}

\vspace{-2.5cm}

\begin{flushright} 
{\small
HISKP-TH-09/22\\
FZJ-IKP-TH-2009-20\\
}
\end{flushright}

\vspace{1.6cm}

\begin{center}
{\huge\bf Frontiers in Nuclear Physics}
\\[1cm]
Symposium in honor of Walter Gl\"ockle's 70$^{\rm th}$ birthday\\[0.15cm]
Bad Honnef, Germany, June 18 - 20, 2009\\[1cm]
{\bf Evgeny Epelbaum}$^{1,2}$ 
and 
{\bf Ulf-G. Mei\ss ner}$^{2,1,3}$ 
\\[0.3cm]
$^1${Forschungszentrum J\"ulich, Institut f\"ur Kernphysik (Theorie)\\ 
and J\"ulich Center for Hadron Physics, D-52425 J\"ulich, Germany}\\[0.3cm]
$^2${Universit\"at Bonn, Helmholtz-Institut f\"ur Strahlen- und Kernphysik (Theorie)\\ 
and Bethe Center for Theoretical Physics, D-53115 Bonn, Germany}\\[0.3cm]
$^3${Forschungszentrum J\"ulich, Institute for Advanced Simulation\\ 
D-52425 J\"ulich, Germany}\\[1cm]
{\large ABSTRACT}
\end{center}
These are the proceedings of the symposium on ``Frontiers in
Nuclear Physics'' held at the Physikzentrum Bad Honnef from June 18 
to 20, 2009. The workshop concentrated on recent advances in the understanding
of the nuclear forces, few-nucleon systems and related topics.
Included are a short contribution per talk.

\vfill

\noindent\rule{6cm}{0.3pt}\\
\footnotesize{$^*$ 
This symposium was supported in parts by funds provided from the
Helmholtz Association to the young investigator group
``Few-Nucleon Systems in Chiral Effective Field Theory'' (grant  VH-NG-222)
and through the virtual institute ``Spin and strong QCD'' (grant VH-VI-231).
This work was further supported by the DFG (SFB/TR 16 ``Subnuclear Structure
of Matter'') and by the  EU-Research Infrastructure Integrating Activity
 ``Study of Strongly Interacting Matter'' (HadronPhysics2, grant n. 227431)
under the Seventh Framework Program of the EU and by BMBF (research grant 06BN411).
}

\end{titlepage}

\section{Introduction}

On January 8 of 2009, Walter Gl\"ockle turned 70. Over many decades, his
work had a lasting impact on the theoretical and experimental investigations
of few-nucleon systems, on the theory of nuclear forces, relativistic
quantum mechanics and most recently, Coulomb effects in few-nucleon
systems. We thus found it appropriate to organize a symposium in honor
of these achievements and to review the state-of-the-art in these fields
and related areas. The meeting was organized at the Physikzentrum
Bad Honnef from June 18-20, 2009, with financial support from the 
the Virtual Institute on ``Spin and Strong QCD'', the Network WP4 (QCDnet)
of the HadronPhysics2 project of the seventh framework of the EU and
the young investigator group ``Few-Nucleon Systems in Chiral Effective Field Theory''
of the Helmholtz Association. The meeting had 35 participants whose names, 
institutes and email addresses
are listed below. 17 of them presented results in half hour presentations.
A short description of the contents of each talk and a list of the most relevant
references can be found below. We felt that this was more appropriate a 
framework than full-fledged proceedings.
Most results are or will soon be published and available on the archives,
so this way we can achieve speedy publication and avoid duplication of
results in the archives.

\bigskip

\noindent
Below follows first the program, then the list of participants 
followed by the abstracts of the talks. All talks 
can also be obtained from the workshop web-site 

\medskip

\centerline{{\tt http://www.itkp.uni-bonn.de/~{\large$\tilde{}$}epelbaum/bh09}~~.}

\bigskip

\noindent
We would like to thank the staff of the Physikzentrum, in particular
Victor Gomer, for the excellent organization of the workshop and all participants 
for their valuable contributions.

\vspace{1.2cm}

\hfill Evgeny Epelbaum and Ulf-G. Mei\ss ner 

\vfill \eject

\section{Program}

\vspace{-0.1cm}

\begin{tabbing}
xx:xx \= A very very very long name \= \kill
{\bf Thursday, June 18, 2009}\\[3mm]
19:00 \> \> {\em Workshop dinner} \\[3mm]
{\bf Friday, June 19, 2009}\\[3mm]
{\em Morning session,  chair: Ulf-G. Mei{\ss}ner}  \\
9:00 \>    U.-G. Mei{\ss}ner     \>  Opening remarks \\
9:10\>        A.~Kievsky (Pisa)  \> 
   Three-nucleon forces: a comparative study   \\
9:45 \>
        J.~Golak (Krakow) \>
  Two-pion exchange currents in the\\
\>\> photodisintegration of the deuteron\\
10:20\>        A.~Nogga (J\"ulich)  \> 
   On the action of four-nucleon forces in 4He \\
10:55\>\> {\em Coffee}\\
11:15 \>
        D.~Lee (Raleigh)\>         
 Effective field theory on a lattice\\
11:50 \>
        J.~Haidenbauer (J\"ulich)  \>
  Aspects of the hyperon-nucleon interaction\\
11:50 \>
        D.~Phillips (Athens, Ohio)  \>
  Compton scattering on Helium-3 \\
13:00\>\> {\em Lunch}\\
{\em Afternoon session, chair: Ulf-G. Mei{\ss}ner} \\
14:30 \> 
        N.~Kaiser (M\"unchen) \>      
 Chiral three-nucleon interaction and 14C beta decay\\
15:05 \>
        H.-W.~Hammer (Bonn) \>      
 Few-body physics with resonant interactions \\
15:40 \>
        F.-K.~Guo (J\"ulich)  \>
  Heavy meson hadronic molecules \\
16:15\>\> {\em Coffee}\\
16:45 \>
         B.~Kubis (Bonn) \>
 Cusps  in $K\to 3\pi$ decays\\
17:20 \>
         W.~Plessas  (Graz) \>
 Baryons as relativistic three-quark systems\\
17:55 \>
         N.~Kalantar-Nayestanaki  \>
 What have we learned about three-nucleon systems\\
\> (Groningen) \> at intermediate energies?\\
18:30 \>
         R.~Beck (Bonn) \>
 Recent results in meson photoproduction at ELSA\\
19:05\>\>{\em End of Session}\\[3mm]
{\bf  Saturday, June 20, 2009}\\[3mm]
{\it  Morning Session, chair: Evgeny Epelbaum}\\
09:30 \>
        Ch.~Elster (Athens, Ohio)\>   
 Faddeev calculations in three dimensions \\
10:05\>
        W.~Polyzou (Iowa) \>
Euclidean formulation of relativistic quantum mechanics\\
10:40 \>\> {\em Coffee}\\
11:15\>
        H.~Witala (Krakow)   \> 
     A novel approach to include the pp Coulomb force\\
\>\> into the 3N Faddeev calculations\\
11:35\>
        H.~Kamada (Kitakyushu)  \>  
        Effects of the $\pi\rho$ exchange three-body force\\
\> \> proton-deuteron scattering\\
12:10\> W. Gl\"ockle \> A short look back\\
12:20 \>  E. Epelbaum    \>  Closing remarks \\
12:25\>\> {\em End of session}\\
12:30\>\>{\em Lunch and end of the Symposium} \\
\end{tabbing}

\section{Participants and their email }


\begin{tabbing}
A very long namexxxxx\=a very long institutexxxxxx\=email\kill
R. Beck\> Univ. Bonn\>beck@hiskp.uni-bonn.de\\
P. Bruns\> Univ. Bonn\>bruns@hiskp.uni-bonn.de\\
S.B. Bour\> Univ. Bonn\>bour@hiskp.uni-bonn.de\\
Ch. Elster\>Ohio Univ. (Athens, USA)\>elster@ohiou.edu\\
E. Epelbaum\>FZ J\"ulich \& Univ. Bonn\>e.epelbaum@fz-juelich.de\\
W. Gl\"ockle\> Ruhr-Univ. Bochum\>walter.gloeckle@tp2.ruhr-uni-bochum.de\\
J. Golak\>Univ. Cracow (Poland)\>ufgolak@cyf-kr.edu.pl\\
F.-K. Guo\>FZ J\"ulich\>f.k.guo@fz-juelich.de\\
J. Haidenbauer\>FZ J\"ulich\>j.haidenbauer@fz-juelich.de\\
H.-W. Hammer\> Univ. Bonn\>hammer@hiskp.uni-bonn.de\\
R. Higa\>Univ. Bonn\>higa@hiskp.uni-bonn.de\\
M. Hoferichter\>  Univ. Bonn\>mhofer@uni-bonn.de\\
H. Kamada \>Kyushu Inst.Tech.(Japan) \>kamada@mns.kyutech.ac.jp \\
N. Kaiser\> TU M\"unchen  \>nkaiser@ph.tum.de \\
N. Kalantar\> KVI (Groningen, Netherlands)\>\\
\>\> nasser@kvi.nl\\
A. Kievsky\> Pisa Univ. (Italy) \>alejandro.kievsky@pi.infn.it\\
S. K\"onig\>Univ. Bonn\>sekoenig@hiskp.uni-bonn.de\\
H. Krebs\> FZ J\"ulich \& Univ. Bonn\>hkrebs@hiskp.uni-bonn.de\\
S. Krewald\> FZ J\"ulich\>s.krewald@fz-juelich.de\\
S. K\"olling\> FZ J\"ulich\>s.koelling@fz-juelich.de\\
B. Kubis\>  Univ. Bonn\>kubis@hiskp.uni-bonn.de\\
D. Lee\> North Carolina State \>djlee3@unity.ncsu.edu\\
\> Univ. (Raleigh, USA) \>\\
M. Lenkewitz\>  Univ. Bonn\>lenkewitz@hiskp.uni-bonn.de\\
M. Mai\>  Univ. Bonn\>mai@hiskp.uni-bonn.de\\
U.-G. Mei{\ss}ner\> Univ. Bonn \& FZ J\"ulich\> meissner@hiskp.uni-bonn.de\\
D. Minossi\>FZ J\"ulich\>d.minossi@fz-juelich.de\\
A. Nogga\>FZ J\"ulich\>a.nogga@fz-juelich.de\\
D. Phillips\> Ohio Univ. (Athens, USA)\> phillips@phy.ohiou.edu\\
W. Plessas\> Graz Univ. (Austria)\> willibald.plessas@uni-graz.at\\
W. Polyzou\> Iowa Univ. (USA)\> polyzou@gelfand.physics.uiowa.edu\\
B. Schoch\> Univ. Bonn\>schoch@physik.uni-bonn.de\\
D. Tolentino\> Univ. Bonn\>tolention@hiskp.uni-bonn.de\\
H. Witala\>Univ. Cracow (Poland)\>ufwitala@cyf-kr.edu.pl\\
A. Wirzba\>FZ J\"ulich\>a.wirzba@fz-juelich.de\\
W. von Witsch\> Univ. Bonn\> vwitsch@hiskp.uni-bonn.de\\
\end{tabbing}

\newabstract 

\begin{center}
{\large\bf Three-nucleon Forces: A Comparative Study}\\[0.5cm]
{\bf A. Kievsky}$^1$, M. Viviani$^1$, L. Girlanda$^{1,2}$ and
L.E. Marcucci$^{1,2}$ \\[0.3cm]
$^1$Istituto Nazionale di Fisica Nucleare, Via Buonarroti 2, 
56100 Pisa, Italy \\[0.3cm]
$^2$Dipartimento di Fisica, Universita' di Pisa, Via Buonarroti 2, 
56100 Pisa, Italy \\[0.3cm]
\end{center}

The use of realistic NN potentials in the description
of the three- and four-nucleon systems gives a
$\chi^2$ per datum much larger than 1 (see for example Ref.\cite{kiev01}).
In order to improve this situation, different three-nucleon
force (TNF) models have been introduced as
the Tucson-Melbourne (TM) and the
Urbana IX (URIX) potentials.
More recently, TNF models have been derived
based on chiral effective field theory at next-to-next-to-leading order.
A local version of this interaction (hereafter referred as N2LO)
can be found in Ref.~\cite{N2LO}.
All these models contain a certain number
of parameters that fix the strength of the different terms. It is a common
practice to determine these parameters from
the three- and four-nucleon binding energies. 

In Table~\ref{tb:table1} we report
the triton and $^4$He binding energies, and the doublet
$n-d$ scattering length $^2a_{nd}$. These results were obtained using the
AV18 or the N3LO-Idaho two-nucleon potentials together with the
AV18+URIX, AV18+TM' and N3LO-Idaho+N2LO TNF models.
The results are compared to the experimental values also reported
in the table. 
\begin{table}[b]
\caption{The triton and $^4$He binding energies (in MeV),
and  doublet scattering length $^2a_{nd}$ (in fm)
calculated using the AV18 and the N3LO-Idaho
two-nucleon potentials, and
the AV18+URIX, AV18+TM' and N3LO-Idaho+N2LO two- and three-nucleon
interactions.}
\label{tb:table1}
\vspace{0.12cm}
\begin{tabular}{@{}llll}
\hline
Potential & $B$($^3$H) & $B$($^4$He) & $^2a_{nd}$ \cr
\hline
AV18            & 7.624    & 24.22   & 1.258 \cr
N3LO-Idaho      & 7.854    & 25.38   & 1.100 \cr
AV18+TM'        & 8.440    & 28.31   & 0.623 \cr
AV18+URIX       & 8.479    & 28.48   & 0.578 \cr
N3LO-Idaho+N2LO & 8.474    & 28.37   & 0.675 \cr
\hline
Exp.            & 8.482    & 28.30   & 0.645$\pm$0.003$\pm$0.007 \cr
\hline
\end{tabular}
\end{table}
\noindent
From the table we observe that only the results obtained
using an interaction model that includes a TNF are close to the
corresponding experimental values. However
the predictions are not in complete agreement with the experimental values. 
In the following we would like to discuss possible modifications to
the TNF models in order to improve the description of these three
quantities. As a two body potential we choose the AV18 interaction.
The TNF models are summed to this interaction and,
in the case of the TM potential, the constants $b$, $d$ 
and the cutoff $\Lambda$ have been varied. In the case of the URIX
potential the constant $A_{2\pi}$ and $U_0$ have been varied. Moreover
the constant $D_{2\pi}$ in front of the anticommutator term and
originally fixed to $D_{2\pi}=(1/4) A_{2\pi}$, was allowed
to vary independently of the value of $A_{2\pi}$. Finally,
for the N2LO potential the constants $c_3$, $c_4$, $c_D$ and $c_E$ have
been varied. As an example, in Fig.1 the values of $D_{2\pi}$
in function of $A_{2\pi}$ are shown as well as the doublet scattering
length and the $^4$He binding energy for the AV18+URIX potential. 
In each point the value of $U_0$, which fixes the strength of the
repulsive term, has been fixed to describe the triton binding energy. 
The values
corresponding to the original potential are indicated by crosses in
the figure.
 
\begin{figure}
\begin{center}
    \includegraphics[width=5cm,angle=270]{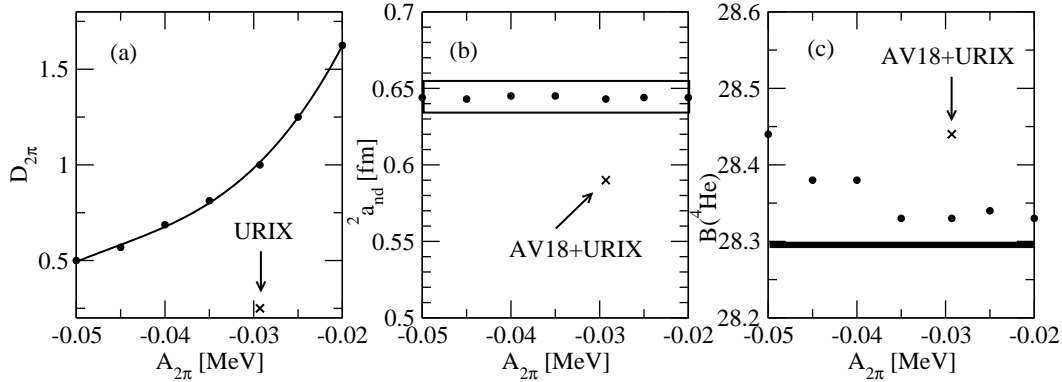}
\end{center}
\caption{Values of $D_{2\pi}$, the doublet scattering length and the
$^4$He binding energy, as a function of $A_{2\pi}$ for the URIX
potential. The crosses correspond to the original values} 
\end{figure}

Similar modifications were performed for the TM and N2LO. A detailed
description of this study for the three TNF potentials under
consideration is in progress~\cite{kiev09}. The main conclusions
are the following. The TM potential, as originally defined, does not
include a repulsive term. Accordingly, a simultaneous description
of $B(^3{\rm H})$, $B(^4{\rm He})$ and $^2a_{nd}$ can be obtained with
unrealistic values of the parameters $b$, $d$ and $\Lambda$. The
introduction of a $c_E$--term helps to obtain a much better
agreement. In the case of the N2LO potential it is possible to obtain
a good agreement with a small change of the original values. The next
step is to analyze the new parametrizations of the potentials in the
description of $N-d$ and $N-^3$He scattering.

\newabstract 

\begin{center}
{\large\bf Two-Pion Exchange Currents in the Photodisintegration of the Deuteron}\\[0.5cm]
{\bf Jacek Golak} and Dagmara Rozp{\e}dzik\\[0.3cm]
M. Smoluchowski Institute of Physics, Jagiellonian University,\\
PL-30059 Krak\'ow, Poland\\[0.3cm]
\end{center}

Chiral effective field theory (ChEFT) is a modern framework 
to analyze properties of few-nucleon systems at low energies
\cite{EE}. It is based on the most general effective
Lagrangian for pions and nucleons consistent with the chiral
symmetry of QCD. For energies below the pion-production 
threshold it is possible to eliminate the pionic degrees 
of freedom and derive nuclear potentials and 
nuclear current operators. This is very important because, despite
a lot of experience gained in the past, the consistence 
between two-nucleon forces, many-nucleon forces and 
corresponding current operators has been not yet achieved.

In this presentation we consider recently derived two-pion
exchange (TPE) contributions to the nuclear current operator~\cite{SiE}.
These operators do not contain any free parameters. 
Due to their isospin structures they do not contribute 
to elastic electron scattering off the deuteron. We thus study 
their role in the deuteron photodisintegration reaction. 
We show how partial wave decomposition for these operator is 
performed using the {\em Mathematica} software and parallel 
computing techniques. We address also a problem of numerical 
stability of some scalar functions which appear in TPE 
contributions. Finally, using a two-nucleon (2N) current operator
containing the single-nucleon, lowest order one-pion exchange 
and TPE contributions, we show predictions for a number 
of observables at photon energy $E_\gamma$= 60 and 120 MeV. 
The bound and scattering states are calculated with 
five different chiral N2LO 2N potentials which results 
in the so-called bands for the predicted results.

For some observables the widths of the bands and their vicinity
to the reference predictions based on the AV18 2N potential
and the current operator (partly) consisted with this force
indicate that the missing contributions in the chiral framework
(like contact currents) do not play a big role.

\newabstract 

\begin{center}
{\large\bf 	On the action of four-nucleon forces in $^{4}$He}\\[0.5cm]
{\bf Andreas Nogga}\\[0.3cm]
Forschungszentrum J\"ulich, Institut f\"ur Kernphysik, Institute 
for Advanced Simulation and J\"ulich Center for Hadron Physics, D-52425 J\"ulich, Germany \\[0.3cm]
\end{center}

Due to the rather good description of {\it ab-initio} calculations for light
nuclei, it is generally expected that the contribution of four-nucleon 
forces (4NF) is insignificant. However, since small deviation of the
predictions remain, it is timely to study the possible impact of the 4NF 
more stringently. In this talk, we present the results of such a study for 
$^4$He within the framework of chiral perturbation theory, where NN, 3N and 
4N forces are formulated consistently on the same footing. 

It was recently shown that the leading chiral 4NF is completely determined 
by parameters of the leading NN interaction
\cite{epelbaum:2006}\cite{epelbaum:2007}. 
Based on this formulation, we have performed a first study of the 4NF 
contribution  to the binding energy of $^4$He \cite{rozpedzik:2006}, where 
we could not take into account the full complexity of the $^4$He wave
function. Due to this, our results especially for the short range 4NF's 
were not reliably enough for final conclusions on the size of 4NF's. 
 
In this talk, we presented for the first time the results of complete 
calculations of the 4NF contribution to the $^4$He binding energy. 
We estimate the 4NF perturbatively. For the pertinent expectation values, 
we devised a Monte Carlo method suitable for calculations in 
momentum space. 
We found that the 4NF contributions depend on the cutoff and order of the 
chiral interactions. They are however in line with power counting estimates. 
Individual classes of diagrams of the 4NF contribute of the order of 1 MeV 
to the binding energy of $^4$He. 
We also found that  attractive and  repulsive classes of diagrams 
cancel each other in parts, so that the net contribution mostly is below 500 
MeV for $^4$He. Since such a cancellation could be less prominent in nuclei 
other $^4$He, we expect a natural contribution of the 4NF of 200-300 keV per 
nucleon. Based on these results, the 4NF will not be negligible in high 
precision nuclear structure calculations. 
The results presented will be published in \cite{rozpedzik:2009}.

\newabstract 

\begin{center}
{\large \textbf{Effective field theory on a lattice}}\\[0.5cm]
{Evgeny Epelbaum}$^{1,2}$, {Hermann~Krebs}$^{2,1}${, \textbf{Dean~Lee}}$%
^{3,2}${, }Ulf-G. Mei\ss ner$^{2,1,4}$\\[0.3cm]
$^{1}$Institut f\"{u}r Kernphysik (IKP-3) and J\"{u}lich Center for Hadron
Physics,

Forschungszentrum J\"{u}lich, D-52425 J\"{u}lich, Germany\\[0.3cm]
$^{2}$Helmholtz-Institut f\"{u}r Strahlen- und Kernphysik (Theorie)

and Bethe Center for Theoretical Physics, \linebreak Universit\"{a}t Bonn,
D-53115 Bonn, Germany \\[0.3cm]

$^{3}$Department of Physics, North Carolina State University, 

Raleigh, NC 27695, USA\\[0.3cm]

$^{4}$Institute for Advanced Simulations (IAS), 

Forschungszentrum J\"{u}lich, D-52425 J\"{u}lich, Germany
\end{center}

Lattice effective field theory combines the theoretical framework of
effective field theory with numerical lattice methods. \ This approach has
been used to simulate nuclear matter \cite{Muller:1999cp} and neutron matter 
\cite{Lee:2004qd}\cite{Lee:2004si}\cite{Abe:2007fe}. \ It has also been used to study
light nuclei in pionless effective field theory \cite{Borasoy:2005yc} and
chiral effective field theory at leading order \cite{Borasoy:2006qn}. \ A
review of the literature in lattice effective field theory can be found in
Ref.~\cite{Lee:2008fa}.

Recently calculations at next-to-leading order in chiral effective field
theory have been carried out for the ground state of dilute neutron matter 
\cite{Borasoy:2007vi}\cite{Borasoy:2007vk}\cite{Epelbaum:2008vj}. \ With auxiliary
fields and Euclidean-time projection Monte Carlo, the ground state energy
for $8$, $12,$ and $16$ neutrons in a periodic cube was calculated for a
density range from 2\% to 10\% of normal nuclear density.

Another recent study considered low-energy protons and neutrons on the
lattice at next-to-next-to-leading order in chiral effective field theory 
\cite{Epelbaum:2009zs}. \ Three-body interactions first appear at this
order, and several methods were considered for determining three-body
interaction coefficients on the lattice. \ The energy of the triton was
calculated as well as low-energy neutron-deuteron scattering in the
spin-doublet and spin-quartet channels using\ L\"{u}scher's finite volume
method \cite{Luscher:1985dn}. \ In the four-nucleon system the energy of the 
$\alpha $-particle was computed using auxiliary fields and Euclidean-time
projection Monte Carlo.

\newabstract 

\begin{center}
{\large\bf Aspects of the hyperon-nucleon interaction}\\[0.5cm]
{\bf Johann Haidenbauer}$^1$ and Ulf-G. Mei\ss ner$^{1,2}$\\[0.3cm]
$^1$Institut f\"ur Kernphysik, Institute for Advanced Simulation and
J\"ulich Center for Hadron Physics, Forschungszentrum J\"ulich,
D-52425 J\"ulich, Germany \\[0.3cm]
$^2$Helmholtz-Institut f{\"u}r Strahlen- und Kernphysik (Theorie) and Bethe
Center for Theoretical Physics,
Universit{\"a}t Bonn, D-53115 Bonn, Germany.\\[0.3cm]
\end{center}

The derivation of nuclear forces within chiral effective field theory
(EFT) has been pursued extensively since the work of Weinberg \cite{Wei90}. 
The most recent publications in this direction demonstrate that the nucleon-nucleon 
($NN$) interaction can be described to a high precision in chiral 
EFT \cite{Entem}\cite{EE}. For reviews we refer the reader to \cite{Rev1}\cite{Rev2}\cite{Rev3}. 

The situation is different for baryon-baryon systems with strangeness.
Only very few studies exist for 
the strangeness $S=-1$ sector, i.e. for the hyperon-nucleon ($YN$) 
interaction ($Y = \Lambda, \ \Sigma$) \cite{YN1}\cite{YN2}\cite{YN3}. 
The strangeness $S=-2$, $-3$, and $-4$ channels have not been 
considered within chiral EFT at all.  

In my talk I present results from our ongoing investigation of 
the baryon-baryon ($BB$) interaction in the strangeness $S=-1$, $-2$, $-3$, 
and $-4$ channels, performed 
within the framework of chiral EFT \cite{Pol1}\cite{Pol2}\cite{Haidenbauer}\cite{Hai}. 
At leading order (LO) in the power counting   
the $BB$ interactions consist of four-baryon contact terms without
derivatives and of one-pseudoscalar-meson exchanges, analogous to the
$NN$ potential of \cite{EE}. The potentials are derived using $SU(3)$
flavor symmetry constraints. Then there are in total only six independent
contact terms whose parameters, the low-energy constants (LECs), need 
to be determined by a fit to data. $SU(3)$ symmetry also interrelates the
coupling constants at the various (pseudoscalar) meson-baryon-baryon
vertices \cite{Pol1}. The reaction amplitudes are obtained
by solving a (single or coupled channels) Lippmann-Schwinger equation 
for the LO potential. 
We use an exponential regulator function to regularize the
potential and apply cutoffs in the range between 550 and 700 MeV,
cf. Ref. \cite{Pol1}\cite{Haidenbauer} for details. 

Five of the six contact terms appear in the strangeness $S=-1$ $BB$ interaction
and can be fixed by a fit to low-energy $\Lambda N$ and
$\Sigma N$ scattering data. It turned out that already at LO in 
chiral EFT a description of the available 35 $YN$ data can be achieved 
that is as good as the one for conventional meson-exchange models \cite{Pol1}.
Furthermore, also the binding energies of the light
hypernuclei are predicted well within chiral EFT \cite{Nogga}\cite{Haidenbauer}.

The additional sixth contact term appears only in the $\Xi N$ and $YY$ systems and
can only be determined in the $S=-2$ sector. Adopting natural
values for this additional contact term, a moderately attractive $\Lambda \Lambda$
interaction is obtained - in line with recent empirical information on doubly
strange hypernuclei \cite{Pol2}. 
Furthermore, we could show that the chiral EFT predictions 
are consistent with the recently deduced doubly strange scattering cross sections. 

With regard to the strangeness $S=-3$ and $-4$ sectors
the occurring five contact terms are the same as those that were already fixed 
from our study of the $YN$ interaction.
Thus genuine predictions can be made for the baryon-baryon interactions
in those channels based on chiral EFT and the assumed ${\rm SU(3)}_{\rm f}$ 
symmetry \cite{Hai}.

\newabstract 

\begin{center}
{\large\bf Compton scattering on Helium-3}\\[0.5cm]
{\bf Daniel R. Phillips}$^{1,2}$\\[0.3cm]
$^1$Department of Physics and Astronomy and Institute for Nuclear\\ and Particle Physics,
Ohio University, Athens, OH 45701, USA\\[0.3cm]
$^2$HISKP (Theorie), Universit\"at Bonn, D-53115 Bonn, Germany.\\[0.3cm]
\end{center}

There has been much recent experimental and theoretical work regarding neutron electromagnetic polarizabilities. These quantities are fundamental properties of the neutron, and contain interesting information on the different mechanisms that contribute to neutron structure. (See Ref.~\cite{Ph09} for a recent overview.) Until a couple of years ago most attention was focused on experiments that used Compton scattering on the deuterium nucleus (elastic or breakup) to obtain constraints on neutron polarizabilities. A series of experiments (see Ref.~\cite{Ph09} for details) has allowed the extraction of useful information on the spin-independent polarizabilities, $\alpha^{(n)}$ and $\beta^{(n)}$. The classic chiral-perturbation-theory predictions for these quantities~\cite{Be91} seem to be in fairly good agreement with these extractions.

The four neutron spin polarizabilities  (here denoted $\gamma_1^{(n)}$--$\gamma_4^{(n)}$)  parameterize the $O(\omega^3)$ response of the neutron to applied electric and magnetic fields, and are also calculable in $\chi$PT. As with $\alpha^{(n)}$ and $\beta^{(n)}$, their leading behaviour is given by constants of QCD: $g_A$, $f_\pi$, and $m_\pi$. But higher-order effects in the spin polarizabilities are known to be sizeable~\cite{spinpols}. The extraction of information on $\gamma_1^{(n)}$--$\gamma_4^{(n)}$ will therefore teach us about the interplay of different effects in neutron electromagnetic structure.
 
Recently we completed the first calculation of Compton scattering from the Helium-3 nucleus~\cite{Ch07}. It showed that polarized Helium-3 targets provide an excellent opportunity to obtain information on the $\gamma_i^{(n)}$. There are, at present, only two constraints on combinations of these quantities, so such experiments---which are within the projected capabilities of the HI$\gamma$S facility~\cite{higsreview}---will significantly advance our knowledge of neutron structure. 

In the low-photon-energy region, $\omega \sim m_\pi^2/M$, with $M$ the nucleon mass, a correct treatment of $\gamma {}^3$He scattering requires as input (1) a three-nucleon current operator that is consistent with the potential employed to bind the nucleus and (2) the fully-interacting Green's function of the three-nucleon state. In the case of the $A=2$ system it has been shown that such a treatment produces the correct Thomson limit for Compton scattering from the nucleus~\cite{Hi05}. Carrying out an analogous treatment in the $A=3$ case is an important future step.

But, observables are much more sensitive to polarizabilities, and especially to spin polarizabilities, at photon energies of order 100 MeV. In this regime the three-nucleon Green's function can be evaluated in perturbation theory. This statement is contained in Weinberg's power counting for the $\gamma NNN \rightarrow \gamma NNN$ operator. This operator has been constructed up to $O(e^2 Q)$ [NLO] in Ref.~\cite{Ch07}, and includes the same mechanisms that were demonstrated to be important in describing Compton scattering from deuterium in this energy range~\cite{Be99}. 

The resulting calculations of $\gamma {}^3$He scattering matrix elements predict cross sections that are appreciably larger than those for deuterium in the same energy range. There is also a larger  absolute sensitivity to $\alpha^{(n)}$ and $\beta^{(n)}$.  Meanwhile the results for the double-polarization observables $\Sigma_z$ and $\Sigma_x$ (asymmetries for circularly-polarized photons on a longitudinally and transversely polarized target) are quite close to those predicted for a free neutron. Our calculation, which includes the full three-body wave function, as well as two-body currents, shows that the dominant effect in these asymmetries is Compton scattering from the unpaired neutron in the polarized ${}^3$He ground state. In consequence, $\Sigma_x$ and $\Sigma_z$ show significant sensitivity to novel combinations of the $\gamma_i^{(n)}$. 

The calculation of Ref.~\cite{Ch07} is a first step towards understanding $\gamma {}^3$He scattering. Several improvements are necessary if a precision extraction of neutron spin polarizabilities from data is sought. This work does, though, open up a new avenue through which information on neutron spin-structure can be obtained. The Bochum-Cracow group, led by Walter Gl\"ockle, played a key role in the use of  ${}^3$He targets to measure neutron form factors (see, e.g., Ref.~\cite{Wu00}). This expertise, used in concert with consistent chiral expansions for interactions and current operators, can, when applied to precise experimental data, lead to similar success in determining the neutron Compton amplitude.

\newabstract 

\begin{center}
{\large\bf  Chiral three-nucleon interaction and $^{14}$C beta decay}\\[0.5cm]
J.W. Holt, {\bf N. Kaiser} and W. Weise\\[0.3cm]
Physik Department T39, Technische Universit\"at M\"unchen,\\
D-85747 Garching, Germany.\\[0.3cm]
\end{center}

The anomalously long lifetime of $^{14}$C, which makes possible 
the radiocarbon dating method, has long been a challenge to nuclear structure 
theory. The transition from the $J_i^\pi = 0^+$, $T_i = 1$ ground state of 
$^{14}$C to the $J_f^\pi = 1^+$, $T_f = 0$ ground state of $^{14}$N is of the 
allowed Gamow-Teller type, yet the known lifetime of $\sim 5730 \pm 30$ years 
is  nearly six orders of magnitude longer than would be expected from typical 
allowed transitions in $p$-shell nuclei. The associated Gamow-Teller
transition matrix element must therefore be accidently small, 
$M_{GT} \simeq 2 \times 10^{-3}$. This feature makes it a sensitive test for
both nuclear interactions and nuclear many-body methods. Recently, it has been 
suggested \cite{holt} that the $^{14}$C beta decay transition matrix element 
should be particularly  sensitive to the density-dependence of the nuclear 
interaction. The study in  ref.\cite{holt} used a medium-dependent
one-boson-exchange interaction modeled with Brown-Rho scaling, where the
masses of the vector mesons decrease due to partial 
restoration of chiral symmetry at  finite density. 

In the present work \cite{c14n3}, we have investigated in detail the role of 
density-dependent corrections to the nuclear interaction 
generated by the leading-order chiral three-nucleon force. 
Pauli-blocking effects by the nuclear medium  introduce a particular 
density-dependence 
into the long-range one-pion and two-pion exchange interactions and also into
the short-range NN-interaction. With the residual nuclear interaction 
$V_{\rm  low-k} +V_{NN}^{\rm med}$ we perform a highly constrained shell-model 
calculation (including second order terms) for the ground state wavefunctions
of the $^{14}$C and $^{14}$N nuclei. After examining the different contributions 
to  the in-medium NN-interaction $V_{NN}^{\rm med}$ we find that the large
suppression of the $^{14}$C beta decay matrix element comes almost entirely
from  to the short-range component of the chiral three-nucleon interaction
$\sim c_E$.

\newabstract 

\begin{center}
{\large\bf Few-Body Physics with Resonant Interactions}\\[0.5cm]
{\bf H.-W. Hammer}\\[0.3cm]
Helmholtz-Institut f\"ur Strahlen- und Kernphysik (Theorie)\\
and Bethe Center for Theoretical Physics,
 Universit\"at Bonn, 53115 Bonn, Germany\\[0.3cm]
\end{center}

Particles with a large scattering length have universal low-energy properties
that do not depend on the details of their interactions at short distances
\cite{Braaten:2004rn}.
These properties include the existence of a geometric spectrum of three-body
bound states (so-called Efimov trimers) and a discrete scale invariance. In
the four-body sector, a new class of universal tetramer states connected to
the trimer states was predicted and recently observed in few-body loss 
processes in an ultracold gas of Cs atoms \cite{Platter:2004qn}. 
Other applications of this approach include halo nuclei
and shallow hadronic molecules.

First, we discuss the possibility to observe excited
Efimov states in $2n$ halo nuclei \cite{Canham:2008jd}. 
Based on the experimental data,
$^{20}$C is the only halo nucleus candidate to possibly have an Efimov
excited state, with an energy less than 7~keV below the scattering threshold.
Second, we investigate the structure of $^{20}$C and other $2n$ halo nuclei.
In particular, we calculate their matter form factors, radii, and
two-neutron opening angles.

The mass and the likely quantum numbers ($J^{PC}=1^{++}$)
of the $X(3872)$ suggest that it is either a
weakly-bound hadronic ``molecule'' 
or a virtual state of neutral charm mesons. Assuming the $X(3872)$ is
a weakly-bound molecule, 
we calculate the phase shifts and cross section for
scattering of $D^0$ and $D^{*0}$ mesons
and their antiparticles off the
$X(3872)$ in an effective field theory for short-range interactions
\cite{Canham:2009zq}.
It may be possible to extract the scattering  within the
final state interactions
of $B_c$ decays and/or other LHC events.

\newabstract 

\begin{center}
{\large\bf Heavy meson hadronic molecules}\\[0.5cm]
{\bf Feng-Kun Guo}$^1$, C. Hanhart$^{1,2}$, S. Krewald$^{1,2}$  and Ulf-G. Mei\ss ner$^{1,2,3}$\\[0.3cm]
$^1$Institut f\"{u}r Kernphysik and J\"ulich
Center for Hadron Physics,\\
Forschungszentrum J\"{u}lich, D-52425 J\"{u}lich, Germany\\[0.3cm]
$^2$Institute for Advanced Simulation,\\
Forschungszentrum J\"{u}lich, D-52425 J\"{u}lich, Germany\\[0.3cm]
$^3$Helmholtz-Institut f\"ur Strahlen- und Kernphysik and
          Bethe Center for Theoretical Physics, 
Universit\"at Bonn, D-53115 Bonn, Germany\\[0.3cm]
\end{center}

Some hidden or open charmed mesons discovered in recent years were
suggested to have exotic nature other than $c{\bar c}$ or $c{\bar
q}$. Among various exotics, hadronic molecules have specific
features. In this talk, we discuss how to identify heavy meson
hadronic molecules. 1) For a loosely $S$-wave bound state, the
effective coupling constant can be related to the binding
energy~\cite{theorem}, which means the structure information of such
a bound state is hidden in the coupling to its components. 2) A
heavy meson hadronic molecule consisting of a light hadron and a
heavy meson should have spin-multiplet partner, and the mass
splitting should be almost the same as the heavy meson hyperfine
splitting~\cite{Guo:2009id}. 3) If there is an $S$-wave bound state,
one would observe a repulsive-like sign of the $S$-wave scattering
length in the attractive channel using lattice
simulations~\cite{sl}. We give two examples, i.e., the $Y(4660)$ and
the $D_{s0}^*(2317)$.

Relating the effective coupling to the binding energy, we show that
the data of the $Y(4660)$~\cite{y4660} support a $\psi'f_0(980)$
bound state interpretation~\cite{Guo:2008zg}. The enhancement at
about 4630~MeV observed in the
$\Lambda_c^+\Lambda_c^-$~\cite{Pakhlova:2008vn} can also be
understood as the same state once taking into account the
$\Lambda_c^+\Lambda_c^-$ $S$-wave final state
interaction~\cite{lclc}. Heavy quark spin symmetry predicts a
pseudoscalar $\eta_c'f_0(980)$ bound state with a mass of
$4616^{+5}_{-6}~{\rm MeV}$ and width of
$\Gamma(Y_\eta\to\eta_c'\pi\pi)=60\pm30~{\rm MeV}$. It can be
searched for in the $B\to K\eta_c'\pi\pi$ and $B\to
K\Lambda_c^+\Lambda_c^-$. There might already have been a signal of
such a state in the BABAR data of the latter
channel~\cite{Aubert:2007eb}, but the bad statistics prevents us
from making any decisive conclusion~\cite{lclc}.

The scattering between charmed mesons and light pseudoscalar mesons
are studied using unitarized chiral perturbation
theory~\cite{Guo:2008gp}\cite{Guo:2009ct}. The scalar and isoscalar
$D_{s0}^*(2317)$ is generated as a bound state pole in the isospin
limit, and as a pole in the second Riemann sheet when the isospin
breaking is allowed. By studying its pole position and the $S$-wave
$DK$ scattering length in the isoscalar channel, we propose two ways
towards identifying the nature of the $D_{s0}^*(2317)$.
Experimentally, we suggest to measure directly its decay width into
the $D_s^+\pi^0$, which is of order of about 100~keV in the hadronic
molecule picture~\cite{Guo:2008gp}\cite{2317}, much larger than that
obtained assuming a $c{\bar s}$ or a tetraquark structure. We also
suggest to calculate the $DK$ scattering length using lattice
simulations, the result for its absolute value would turn out to be
about 1~fm if there is an $S$-wave $DK$ bound state, and smaller if
it has an elementary $c{\bar s}$ component. Thus the nature of the
$D_{s0}^*(2317)$ could be measured directly in the lattice.

\newabstract 

\begin{center}
{\large\bf Cusps in \boldmath{$K\to3\pi$} decays}\\[0.5cm]
{\bf Bastian Kubis}\\[0.3cm]
   Helmholtz-Institut f\"ur Strahlen- und Kernphysik (Theorie)
   and 
   Bethe Center for Theoretical Physics,
   Universit\"at Bonn, D--53115~Bonn, Germany\\[0.3cm]
\end{center}

It has been pointed out by Cabibbo and Isidori~\cite{Cabibbo}
that the pion mass difference generates a pronounced cusp in $K^+ \to \pi^0\pi^0\pi^+$ decays,
an accurate measurement of which may allow one to determine the combination $a_0-a_2$ of
S-wave $\pi \pi$ scattering lengths to high precision.   
A first analysis of the data taken by the NA48/2 collaboration was performed in Ref.~\cite{Batley}.
In order for this program to be carried out successfully, 
one needs to determine the structure of the cusp with a precision that
matches the experimental accuracy.  

Non-relativistic effective field theory is the appropriate systematic framework 
to analyze the structure of $K\to 3\pi$ amplitudes and their dependence on the $\pi\pi$ 
scattering lengths~\cite{CGKR}, as the non-relativistic Lagrangian directly contains
the parameters of the effective range expansion of the $\pi\pi$ scattering
amplitude.  
This is in contrast to, e.g., chiral perturbation theory, where scattering lengths etc.\ 
are expanded in powers of the pion mass~\cite{Gamiz};
see also Ref.~\cite{Kampf} for an alternative approach.
As the $\pi\pi$ scattering lengths are small, it is useful to perform a combined
expansion in powers of the scattering lengths and a non-relativistic small parameter $\epsilon$.

The power counting is set up such that pion three-momenta are counted as 
$\mathcal{O}(\epsilon)$; the kinetic energies are therefore  
of $\mathcal{O}(\epsilon^2)$, 
and consequently so is the mass difference $M_K-3M_\pi$.
In this way, the non-relativistic region covers the whole decay region.
It is convenient to formulate the non-relativistic approach in a manifestly
Lorentz-invariant/frame-independent manner,
which can be achieved by employing a non-local kinetic-energy Lagrangian for 
the pion fields of the form 
\begin{equation}
 \mathcal{L}_{\rm kin} = \Phi^\dagger (2W)(i\partial_t-W)\Phi ~, \qquad W=\sqrt{M_\pi^2-\triangle} ~.
\end{equation}
Non-relativistic Lagrangians for $K \to 3\pi$ as well as for the $\pi\pi$ interaction 
are given by
\begin{eqnarray}
\mathcal{L}_{K} &=& \frac{G_0}{2} \bigl(K_+^\dagger \Phi_+ (\Phi_0)^2 + h.c. \bigr) 
          +  \frac{H_0}{2} \bigl(K_+^\dagger \Phi_- (\Phi_+)^2 + h.c. \bigr) + \ldots ~, \nonumber\\
\mathcal{L}_{\pi\pi} &=& C_x \bigl( \Phi_-^\dagger \Phi_+^\dagger (\Phi_0)^2 + h.c. \bigr) + \ldots ~,
\end{eqnarray}
where the ellipses include higher-order derivative terms.
The parameters $G_0$, $H_0$, $C_x$, etc.\ have to be determined from a 
simultaneous fit of $K^+ \to \pi^0\pi^0\pi^+$ and $K^+ \to \pi^+\pi^+\pi^-$ amplitudes
to experimental data.
$C_x$ in particular is proportional to $a_0-a_2$ up to isospin breaking corrections.

The non-relativistic representation of the $K \to 3\pi$ amplitudes has been worked out
up to $\mathcal{O}(\epsilon^4, a \epsilon^5, a^2\epsilon^4)$.
This representation is valid to arbitrary orders in the quark masses.
As the approach is based on a Lagrangian framework, constraints from analyticity
and unitarity are automatically obeyed.
Special care has to be taken for the representation of overlapping two-loop graphs,
which have a particularly complicated analytical structure and which,
for certain combinations of pion masses running in the loops, develop 
anomalous thresholds in the decay region.  

The non-relativistic Lagrangian approach has been extended
to include radiative corrections due to real and virtual photons~\cite{Photons}. 
Photons modify the singularity structure at threshold at $\mathcal{O}(\alpha)$,
and therefore can have a significant effect on the scattering length extraction.

Similar cusp effects also occur in other decay channels, 
such as $K_L,\,\eta\to3\pi^0$~\cite{KLeta} or $\eta'\to\eta\pi^0\pi^0$~\cite{etaprime}.
The extent to which they affect the decay spectra however critically depends 
on the relative decay strength into the charged vs.\ the neutral final states, which
turns out to be rather small for the $K_L$ and $\eta$ decays, but more promising
in the case of the $\eta'$.  
Non-analytic effects due to $\pi\eta$ rescattering at the border of the Dalitz plot
in the latter channel cancel and cannot be observed.

\newabstract 

\begin{center}
{\large\bf Baryons as Relativistic Three-Quark Systems}\\[0.5cm]
{\bf Willibald Plessas} \\[0.3cm]
Theoretical Physics, Institute of Physics, University of Graz,\\
Universit\"atsplatz 5, A-8010 Graz, Austria
\end{center}

A selective review of the performance of relativistic constituent-quark models for
low-energy baryon physics is given.  

There has been a vivid development of constituent-quark models with regard to baryon
spectroscopy over the past decade. At present one of the most realistic descriptions of
the (low-lying) spectra of all light, strange, and charmed baryons in a unified framework
is provided by the relativistic constituent-quark model (RCQM) relying on Goldstone-boson
exchange (GBE)
\cite{Glozman:1998ag}\cite{Glantschnig:2005}\cite{Prieling:2006}.
A similar attempt is, e.g., undertaken by the Bonn group
\cite{Loring:2001kx}. The problem of level orderings in the
$N^*$ and $\Lambda^*$ spectra appears to be resolved, while, most prominently, the
$\Lambda$(1405) remains as a notorious problem \cite{Plessas:2003av}.

What concerns the nucleon ground states, many phenomenological evidences are available
from experiments. Certainly, a realistic quark model should be able to describe them. The electromagnetic structures of both the proton and the neutron are well predicted by the
RCQM at low momentum transfers up to $Q^2 \sim 4$ GeV$^2$
\cite{Wagenbrunn:2000es}\cite{Berger:2004yi}, i.e. within a range where the validity of a
constituent-quark model can be expected. In this context it is essential to look at
covariant results, and
relativistic effects are important in any respects, even in low-momentum-transfer
observables such as the electric radii and magnetic moments of the ground-state baryons
\cite{Berger:2004yi}. While the Graz group produced
these covariant results in point-form relativistic quantum mechanics, the Bonn group
arrived at qualitatively similar predictions through a Bethe-Salpeter approach
\cite{Merten:2002nz}. Essentially the same is true with regard to the axial and induced
pseudoscalar form factors, $G_A(Q^2)$ and $G_P(Q^2)$, of the nucleons, which are naturally described by the GBE RCQM in agreement with phenomenology \cite{Glozman:2001zc}.

The situation is not yet so clear-cut with regard to baryon resonances. Only in recent
years, first covariant results have become available from RCQMs for the various strong decay modes of light and strange resonances \cite{Metsch:2003ix}\cite{Melde:2005hy}. In general,
the decay widths contain considerable relativistic effects but turn out to be too small,
hinting to defects in the assumed decay dynamics.
Nevertheless, one has obtained interesting insights into the resonance structures, leading
to a partially new classification into flavor multiplets \cite{Melde:2008yr}.

Most recent results from the RCQM concern the structures of strong meson-baryon interaction
vertices \cite{Melde:2009} and the axial charges of the nucleon and $N^*$ resonances
\cite{Choi:2009}.

In summary, relativistic constituent-quark models, if well done, appear to be capable to
serve as an effective tool for a comprehensive description of the wealth of low-energy
hadronic phenomena on a uniform basis.

\newabstract 

\begin{center}
{\large\bf What have we learned about three-nucleon systems at intermediate 
energies?}\\[0.5cm]
{\bf N. Kalantar-Nayestanaki}\\[0.3cm]
KVI, University of Groningen,\\
Groningen, The Netherlands\\[0.3cm]

\end{center}

Three and four-body systems have been studied in detail at KVI and other laboratories 
around the world in the last few years. 
Two categories of reactions have been chosen to investigate these systems: i)
elastic, transfer and break-up reactions in proton-deuteron and deuteron-deuteron 
scattering in which only hadrons are involved, and ii) proton-deuteron capture reaction involving 
real and virtual photons in the final state. In this presentation, I focus mainly
on the reactions where only hadrons were involved. 

Hadronic reactions excluding photons give a handle on effects 
such as those from three-body forces. Three-body forces are, though small, 
very important in nature. The effect of these forces have far-reaching 
consequences in many fields of physics. Even though a relatively good 
understanding of most phenomena in nuclear physics has been arrived at
by only considering two-nucleon forces, high precision three-nucleon data have
revealed the shortcomings of these forces. In the last few decades, the two-nucleon 
system has been thoroughly investigated both experimentally and theoretically. 
These studies have resulted in modern potentials which describe the bulk of 
the data in a large range of energy. This knowledge can be employed in a 
Faddeev-like framework to calculate scattering observables in three-body systems \cite{gloeckle}. 
In regions and for the reactions in which the effects of Coulomb force are expected 
to be small or can be calculated accurately, and energies are low enough to avoid 
sizable relativistic effects, deviations from experimental data must then be a 
signature of, for instance, three-body force effects. 

At KVI and other laboratories, various combinations of high-precision cross sections, analyzing 
powers and spin-transfer coefficients have been measured at different incident 
proton or deuteron beam energies between 100 and 200 MeV for a large range of 
scattering angles and for all the reactions mentioned above. Calculations 
based on two-body forces only do not describe the data sufficiently. The inclusion of 
three-body forces improve these discrepancies with data significantly. However, there 
are still clear deficiencies in the calculations. Due to the extended data set which has
been made available, one should now be able to develop a similar approach like 
the partial-wave analysis in the two-nucleon sector to see where the possible problems might 
lie. Results of some recent measurements for a number of observables at intermediate energies performed at KVI and RIKEN 
for the elastic scattering 
\cite{Bieber}--\cite{Ramazani}
as well as for the break-up channel 
\cite{Kistryn1}--\cite{Eslami1} 
in proton-deuteron system were presented. 
In addition, preliminary results were shown for the deuteron-deuteron scattering where all possible outgoing
channels were studied. For the four-body system, the theoretical developments are in their
infancy. This makes this field of research very exciting as experimental results which have been obtained
recently will be available to put to test the results of any ab-initio calculations for the four-body system. 

The three-nucleon results presented in the talk have been obtained in a long-term collaboration between KVI and the 
Cracow and Katowice groups. The four-body systems were recently initiated at KVI with the addition of the IUCF group in 
the collaboration.

\newabstract 

\begin{center}
{\large\bf Recent Results in Meson Photoproduction at ELSA}\\[0.5cm]
{\bf R. Beck}\\[0.3cm]
Helmholtz-Institut f\"ur Strahlen- und Kernphysik\\
Universit\"at Bonn, 53115 Bonn, Germany\\[0.3cm]
\end{center}

The leading goal of the experimental program within the CBELSA/TAPS 
collaboration is to perform precise measurements on photoproduction of 
mesons on the nucleon in the mass region up to 2.5~GeV with the explicit
inclusion of polarization degrees of freedom to extract information
on the dynamics of the production process and on the baryon spectrum.
Of crucial importance for a detailed understanding of the baryon spectrum is 
the measurement of single and double polarization observables to reduce 
the existing ambiguities in the partial wave analysis and to increase the 
sensitivity on small resonance contributions. The final goal is of 
course to get as close as possible to a complete data base, which would 
allow for a model-independent partial wave analysis. For a complete data 
base, allowing for this model-independent partial wave analysis, eight carefully 
chosen observables need to be measured \cite{tabakin}.

{\small
\begin{table}[h]
\begin{center}
{
\setlength{\tabcolsep}{0.2em}
\begin{tabular}{|c|c|ccc|ccc|cccc|}\hline
Photon                   & ~        & \multicolumn{3}{c|}{{ Target}} &
\multicolumn{3}{c|}{Recoil} & \multicolumn{4}{c|}{Target--Recoil} \\\hline
~                        & ~        &  ~   &    ~  & ~     & $x'$    & $y'$ &
$z'$
& $x'$      & $x'$       & $z'$       & $z'$ \\
~                        & ~        & $x$  & $y$   & $z$   &    ~    & ~    &
~
& $x $      & $z $       & $x $       & $z $ \\\hline
{ unpolarized}     & \color{green}\boldmath$\sigma$ & $0$  &
{\color{red}\boldmath$T$}
   & $0$   & $0$     & {$P$}  & $0$    & $T_{x'}$  & -$L_{x'}$  &
$T_{z'}$   & $L_{z'}$\\
{ linear }         &\color{green} (-{\boldmath$\Sigma$}) &
{\color{red}\boldmath$H$}
& {\color{red}$($-\boldmath${P})$}&  {\color{blue} (-\boldmath$G$)}  &
$O_{x'}$& { (-$T$)}&$O_{z'}$& (-$L_{z'})$& $(T_{z'})$ &
$($-$L_{x'})$& $($-$T_{x'})$\\
{ circularly }     & $0$      & {\color{red}\boldmath$F$}  & $0$   & {
\color{blue}
(-\boldmath$E$)}  & (-$C_{x'}$)&$0$   &(-$C_{z'}$)& $0$       & $0$        &
$0$        &
$0$\\\hline
\end{tabular}
}
\end{center}
\caption{\label{a1:table_polobs}
{\small Observables in single pseudoscalar meson photoproduction.
The observables in green have been measured in the past, the ones in blue are
presently measured
using a longitudinally polarized or
unpolarized target and a linear, circular or unpolarized photon beam.
Red: Single and double polarization observables accessible
with a transversally polarized target and a polarized or unpolarized beam, not
measuring the recoil polarization.}}
\label{polobs}
\end{table}
}

Table~\ref{polobs} shows the observables accessible in single pseudoscalar 
photoproduction, which have been already measured (green) at ELSA
for $p\pi^0$ \cite{pion} and $p\eta$ \cite{eta}. In blue shown are the
observables that are presently measured with
a longitudinally polarized target and linearly as well as circularly polarized
photons. Preliminary results for the observables G and E in the channels 
$\vec{\gamma}\vec{p} \rightarrow p\pi^0$  and  $\vec{\gamma}\vec{p} 
\rightarrow p\eta$ have been presented at the NSTAR2009 conference~\cite{nstar}.
In red the observables are shown that become accessible using a 
transversally polarized target. 

A complete data base for pseudoscalar meson photoproduction (the 
``complete'' experiment) requires at least eight independent observables to be 
measured. Such complete information is not available at present, however, 
close to thresholds or in the $P_{33}(1232)$ resonance region, where only a 
few partial waves contribute, an almost model-independent analysis can be 
performed. This has been demonstrated in the s- and p-wave determination at 
the $\pi^0$-threshold \cite{schmidt}, where information on the differential cross 
section ($\frac{d\sigma}{d\Omega}$) and the photon beam asymmetry ($\Sigma$) were 
sufficient to determine the four s- and p-wave amplitudes ($E_{0+}, 
M_{1+}, M_{1-}$ and $E_{1+})$. Another nice example is the determination
of the $E2/M1$-ratio of the $P_{33}(1232)$-resonance. Again, precise
data on $\frac{d\sigma}{d\Omega}$ and $\Sigma$ for the reactions $\vec{\gamma}  p 
\rightarrow p \pi^0$ and $\vec{\gamma}  p \rightarrow n \pi^+$ have been
used to determine the isospin 1/2 and 3/2 contributions of the four
s- and p-waves \cite{beck}.
One important constraint in this analysis is the Fermi-Watson theorem. 
The multipole amplitudes $M^{I}_{l\pm}$ are complex functions of 
the c.m. energy $W$. Below the two-pion production threshold, the 
Fermi-Watson theorem allows one to express the phases of the complex 
multipole amplitudes by the corresponding pion-nucleon scattering phase 
shifts. Above the two-pion production threshold a nearly complete data base 
is necessary to constrain the PWA-solutions. The new Crystal Barrel 
experiment at ELSA will make major contributions to the new
meson-photoproduction data base.

\newabstract 

\begin{center}
{\large\bf  Faddeev Calculations in Three Dimensions}\\[0.5cm]

{\bf Ch. Elster}\\[0.3cm]
Institute of Nuclear and Particle Physics and \\
Department of Physics and Astronomy, Ohio University\\ 
Athens, OH 45701, USA\\[0.3cm]
\end{center}

An enormous effort has been made to understand the scattering of three nucleons
in the energy regime below the pion-production threshold\cite{wgphysrep}. However,
if one wants to understand the same reactions in the intermediate energy regime, 
the standard partial wave description, successfully applied at lower energies, is
no longer an adequate numerical scheme due to the proliferation of the number of
partial waves. In addition, a consistent treatment of intermediate energy
reactions requires a Poincar\'e symmetric quantum theory \cite{Wigner39}.
Thus, the intermediate energy regime is a new territory for
few-body calculations, which waits to be explored.

In this work two aspects in this list of challenges are addressed: exact Poincar\'e
invariance and calculations using vector variables instead of partial waves. 
In order to carry out this work a simplification in the underlying force has been
made, namely scalar nuclear forces are employed consisting of a superposition of
an attractive and repulsive Yukawa force such that a two-body bound state at
$E_d$~=~-2.23~MeV is supported. The three-body bound state (including scalar
three-body forces) was computed in Ref.~\cite{Liu:2002gh}. In
Ref.~\cite{Liu:2004tv} the non-relativistic Faddeev equations were
solved directly as function of vector variables for scattering 
up to 1~GeV, and the feasibility as well as numerical reliability of the approach
were established. 

The Faddeev equation, based on a Poincar{\'e} invariant mass
operator has been formulated in detail in~\cite{Lin:2007ck}
and has both kinematical and dynamical
differences with respect to the corresponding non-relativistic
equation. The formulation of the theory is given in a representation of
Poincar\'e invariant quantum mechanics where the interactions are
invariant with respect to kinematic translations and rotations
\cite{Coester65}.  The model Hilbert space is a three-nucleon Hilbert
space (thus not allowing for absorptive processes).  The method
introduces the NN interactions in the unitary
representation of the Poincar\'e group and allows to input  e.g.
high-precision NN interactions in a way
that reproduces the measured two-body observables.
Poincar\'e invariance and $S$-matrix cluster
properties dictate how the two-body interactions must be embedded in
the three-body dynamical generators.  Scattering observables are
calculated using Faddeev equations formulated with the mass Casimir
operator (rest Hamiltonian) constructed from these generators.

To obtain a valid estimate of the size of relativistic effects, it is
important that the interactions employed in the relativistic and
non-relativistic calculations are phase-shift equivalent. We follow
the suggestion by Coester, Piper, and Serduke (CPS) and construct a
phase equivalent interaction from a non-relativistic 2N
interaction~\cite{CPS}\cite{Keister:2005eq} by adding the interaction to the square of the
mass operator.
Thus, differences in the relativistic and non-relativistic
calculations first appear in the three-body calculations.
Those differences are in the choice of kinematic variables (Jacobi
momenta are constructed using Lorentz boosts rather then Galilean
boosts) and in the embedding of the two-body interactions in the
three-body problem, which is a consequence of the non-linear relation
between the two and three-body mass operators. These differences
modify the permutation operators and the off-shell properties of the
kernel of the Faddeev equations. We studied three-body elastic scattering as well
as breakup reactions~\cite{Lin:2007kg}\cite{Lin:2008sy}, and found that especially in
breakup reactions relativistic effects can be quite large (depending on the
configuration) already at energies as low as 500~MeV. We also found that the
Faddeev multiple scattering series converges rather rapidly once the projectile
laboratory energy exceeds 1~GeV.

\newabstract 

\begin{center}
{\large\bf Euclidean formulation of relativistic quantum mechanics}\\[0.5cm]
{\bf W. N. Polyzou} and Philip Kopp  \\[0.3cm]
Department of Physics and Astronomy,\\
Iowa City, IA 52246 \\[0.3cm]
\end{center}
We discuss preliminary work on a formulation of relativistic quantum
mechanics that uses reflection-positive Euclidean Green functions or
generating functionals as phenomenological input.  This work is
motivated by the Euclidean axioms of quantum field theory \cite{os}\cite{fr}.
The key observations are (1) locality is not used to
reconstruct the quantum theory and (2) it is possible to construct
a fully relativistic quantum theory without performing an explicit
analytic continuation.

Hilbert space vectors are represented by wave functionals $A[\phi]$
with inner product
\[
A[\phi] = \sum_{j=1}^{n_a} a_j e^{i \phi (f_j) }
\qquad
\langle A \vert B \rangle :=
\sum_{j,k}^{n_a,n_b}  a_l^* b_k Z[g_g-\Theta f_j]
\]
where $a_j$ are complex constants, $f_j$ are real Schwartz functions on 4
dimensional Euclidean space with positive-time support, $\Theta$ is
the Euclidean time-reflection operator, and $Z[f]$ is the Euclidean
generating functional.  Reflection positivity is the condition that
$\langle A \vert A \rangle \geq 0$.  For $\beta \geq 0$ and
$\mathbf{a} \in \mathbb{R}^3$ we define
\[
T(\beta, \mathbf{a}) A[\phi] := 
\sum_{j=1}^{n_a} a_j e^{i \phi (f_{j,\beta,\mathbf{a}}) }
\qquad 
f_{j,\beta,\mathbf{a}} (\tau, \mathbf{x}) := 
f_{j} (\tau-\beta , \mathbf{x}-\mathbf{a}) . 
\]
The square of the mass operator operating on a wave functional 
$A[\phi ]$ is 
\[
M^2 A[\phi] = \left ({\partial^2 \over \partial \beta^2} +
{\partial^2 \over \partial \mathbf{a}^2} \right )
T(\beta, \mathbf{a}) A[\phi]_{\vert \beta = \mathbf{a}=0}. 
\]
Solutions of the mass eigenvalue problem with eigenvalue $\lambda$ can
be expanded in terms of an orthonormal set of wave functionals
$A_n[\phi]$, $\langle A_n \vert A_m \rangle = \delta_{mn}$: 
\[
\Psi_{\lambda} [\phi] = \sum \alpha_n A_m [\phi] .
\]
Simultaneous eigenstates of mass, linear momentum, spin, and $z$ component of 
spin can be constructed from $\Psi_{\lambda} [\phi]$ using 
\[
\Psi_{\lambda, j, \mathbf{p},\mu }[\phi]=
\int_{SU(2)} dR \int_{\mathbb{R}^3}
{d\mathbf{a} \over (2 \pi )^{3/2}}  
e^{-i \mathbf{p} \cdot R \mathbf{a}}
U(R) T(0, \mathbf {a})\Psi_\lambda [\phi] 
D^{j*}_{\mu j}[R] 
\]
where $U(R)$ rotates the vector arguments of $f_j(\tau,\mathbf{x})$ in
$A[\phi]$.  When $\lambda$ is in the discrete spectrum of $M$,  
$\Psi_{\lambda, j, \mathbf{p},\mu }[\phi]$ is a wave functional for
a single-particle state that necessarily transforms as a mass 
$\lambda$ spin $j$ 
{\it irreducible representation}.

Products of suitably normalized single-particle wave functionals
define mappings from the product of single-particle irreducible
representation spaces of the Poincar\'e group to the model Hilbert
space.  Because these wave functionals create only single particle
states out of the vacuum, their products are Haag-Ruelle injection
operators \cite{jost}\cite{simon} for the two-Hilbert-space formulation 
\cite{simon} of scattering theory.
If we define $\Phi [\phi] := \prod_k \Psi_{\lambda_k, j_k,
\mathbf{p}_k,\mu_k }[\phi]$, $\otimes g_k = \prod g_k
(\mathbf{p}_k,\mu_k)$, and $H_f= \sum_k \sqrt{\lambda_k^2 +
\mathbf{p}_k^2}$, then scattering wave operator can be defined by the
limit
\[
\Omega_{\pm} \vert \otimes g_k  \rangle :=  
\lim_{t \to \pm\infty} e^{iHt} \Phi e^{-iH_ft} \vert \otimes g_k \rangle.
\]
Using the Kato-Birman invariance principle \cite{simon} to replace 
$H$ by  $-e^{-\beta H}$ gives
\[
\Omega_{\pm} \vert \otimes g_k \rangle :=  
\lim_{n \to \pm\infty} e^{-ine^{-\beta H}}
\Phi e^{ine^{-\beta H_f}} \vert \otimes g_k \rangle . 
\]
Since the spectrum of $e^{-\beta H}$ is compact, for large {\it fixed} $n$ 
$e^{-ine^{-\beta H}}$ can be uniformly approximated by a polynomial in 
$e^{-\beta H}$, which is easy to calculate in this framework.
These steps provide a means to construct all single-particle states,
all scattering states, and compute the action of the Poincar\'e group
on all single-particle states and $S$-matrix elements, using only
the Euclidean generating functional as input.
The advantages of this framework are the relative ease with which cluster 
properties can be satisfied, the close relation to the quantum mechanical 
interpretation of quantum field theory,  and the ability to perform 
calculations directly in Euclidean space without analytic continuation. 

We tested the general method for calculating scattering observables
using a solvable quantum mechanical model of the two-nucleon system.
These test calculations, which used narrow wave packets, a large
finite $n$ and a Chebyshev polynomial expansion of $e^{inx}$,
exhibited convergence to the exact transition matrix elements for a range
of relative momenta between about 100 MeV up to 2 GeV.  This success
warrants further investigation of this framework.

This research was supported by the U.S. Department of Energy.

\newabstract 

\begin{center}
{\large\bf A novel approach to include the pp Coulomb force into the 3N
  Faddeev calculations}\\[0.5cm]
{\bf H. Wita{\l}a}$^1$, R. Skibi\'nski$^1$, J. Golak$^1$ 
and W.\ Gl\"ockle$^2$\\[0.3cm]
$^1$M. Smoluchowski Institute of Physics, Jagiellonian
University, PL-30059 Krak\'ow, Poland,\\[0.3cm]
$^2$Institut f\"ur theoretische Physik II,
Ruhr-Universit\"at Bochum, D-44780 Bochum, Germany.\\[0.3cm]
\end{center}

The  long-range nature of the Coulomb force 
prevents the application of the standard techniques developed for
short-range interactions   in  the analysis of nuclear
reactions involving two protons. 
 In \cite{elascoul} we developed a novel approach 
to include the pp Coulomb
force into the momentum space 3N Faddeev calculations. 
 It is based on a standard
 formulation for short range forces and relies on the screening  of the
long-range Coulomb interaction. In order to avoid  all uncertainties
connected with the application of the partial wave expansion, 
 inadequate  when working
with long-range forces, we used directly the 3-dimensional pp screened
Coulomb t-matrix \cite{skib2009}.
We demonstrated the 
feasibility of that approach in case of elastic pd scattering 
using a simple dynamical model for the nuclear part of the interaction. 
 It turned out that the screening limit exists without the need of 
renormalization not only for pd elastic scattering 
observables but for the elastic pd amplitude itself. 

In \cite{elascoul} we extended that  approach to 
 the pd breakup. Again 
 we apply directly the 3-dimensional screened pp Coulomb
t-matrix without relying on a partial wave decomposition. 
 In \cite{elascoul} 
we demonstrated 
that the physical pd elastic  scattering amplitude can be obtained from
the off-shell solutions of the Faddeev equation and has a well
defined screening limit. In
contrast to elastic scattering, where the amplitude itself 
 does not require renormalization, in case of the pd breakup the
on-shell solutions of the Faddeev equation are required. They 
 demand renormalization in the screening limit which can be achieved 
 through renormalization of the pp t-matrices.

\newabstract 

\begin{center}
{\large\bf  Effects of the $\pi\rho$  exchange three-body force in proton-deuteron scattering
  }\\[0.5cm]
{\bf H. Kamada}$^1$, W. Gl\"ockle$^2$, H. Wita\l a$^3$, A. Nogga$^4$, J. Golak$^3$, R. Skibi\'nski$^3$\\[0.3cm]
$^1$Dep. of Physics, Faculty of Engineering, Kyushu Institute of Technology,\\
1-1 Sensuicho Tobata, 804-8550 Kitakyushu, Japan\\[0.3cm]
$^2$Institut f\"ur theoretische Physik II, Ruhr-Universit\"at Bochum, \\
D-44780 Bochum, Germany\\[0.3cm]
$^3$M. Smoluchowski Institute of Physics, Jagiellonian University, \\
PL-30059 Krak\'ow, Poland\\[0.3cm]
$^4$Forschungszentrum J\"ulich, Institut f\"ur Kernphysik and J\"ulich Center for Hadron Physics, D-52425
J\"ulich, Germany\\[0.3cm]

\end{center}
The Tucson-Melbourne three-body force (3BF) has been investigated by calculating triton binding energy \cite{1} 
and three-nucleon continuum \cite{2}. Using the recent partial wave decomposition scheme PWD \cite{4} of 
the 2$\pi$ exchange type of 3BF 
we calculated the elastic observables of the pd elastic scattering in intermediate energies 
to show \cite{3} the significant effects. We apply the same scheme not only to the 2$\pi$ exchange 3BF 
but also to the $\pi\rho$ exchange one. In the pd elastic scattering at 135MeV/u we compare these theoretical 
predictions (cases without 3BF, with 2$\pi$ 3BF and with 2$\pi$+$\pi\rho$ 3BF) to recent data\cite{5} .
Although the effects of $\pi$ $\rho$ exchange 3BF are almost invisible except for some observables ($A_{xz}$ etc.). 
The PWD technique
will be used \cite{6} to the 3BF of N$^3$LO version in the chiral field theory.

\end{document}